\documentclass{iopart}
\usepackage{setstack}
\usepackage{iopams}
\usepackage{epsfig}

\newcommand {\zz}{ZrZn$_2$ }
\newcommand {\ZZ}{ZrZn$_2$ }

\newcommand {\He}{$^3$He }
\newcommand {\I}{\rmi}
\newcommand {\D}{\rmd}
\renewcommand {\vec}{\bi}

\begin{document}

\title{The gap equations for spin singlet and triplet ferromagnetic superconductors}
\author{B J Powell$^{1,2}$, James F
Annett$^1$ and B L Gy\"orffy$^1$}
\address{$^1$ H H Wills Physics Laboratory, University of Bristol, Tyndall Avenue,
Bristol BS8 1TL, UK}
\address{$^2$ Department of Physics,
University of Queensland, Brisbane, Queensland 4072, Australia}

\ead{powell@physics.uq.edu.au}

\begin{abstract}
We derive gap equations for superconductivity in coexistence with
ferromagnetism. We treat singlet states and triplet states with
either equal spin pairing (ESP) or opposite spin pairing (OSP)
states, and study the behaviour of these states as a function of
exchange splitting. For the s-wave singlet state we find that our
gap equations correctly reproduce the Clogston--Chandrasekhar
limiting behaviour and the phase diagram of the
Baltensperger--Sarma equation (excluding the FFLO region). The
singlet superconducting order parameter is shown to be independent
of exchange splitting at zero temperature, as  is assumed in the
derivation of the Clogston--Chandrasekhar limit. P-wave triplet
states of the OSP type, behave similarly to the singlet state as a
function of exchange splitting. On the other hand, ESP triplet
states show a very different behaviour. In particular there is no
Clogston--Chandrasekhar limiting and the superconducting critical
temperature, $T_{C}$, is actually increased by exchange splitting.
\end{abstract}

\submitto{\JPA}


\section{Introduction}

The recent discovery of the coexistence of ferromagnetism and
superconductivity in UGe$_2$ \cite{Saxena} URhGe \cite{URhGe} and
ZrZn$_2$ \cite{Pfleiderer} has led to renewed interest in the
relationship between ferromagnetism and superconductivity. By
contrast, the relationship between antiferromagnetism and
superconductivity has been more thoroughly studied \cite{Buzdin},
since it is relevant to many compounds, such as the cuprates
\cite{Bulut}, borocarbides \cite{Eskildsen}, heavy Fermion
superconductors \cite{Mathur} and the layered organic
superconductors \cite{Ross_review}.

Particular interest has been focused on superconductivity on the
border of a magnetic phase and in particular in the vicinity of a
quantum critical point (QCP). This is observed experimentally in
cuprates, several heavy Fermion systems, layered organics and UGe$_2$.
It is also
thought that URhGe$_2$ may be essentially similar to UGe$_2$
but under the
influence of \lq chemical pressure' \cite{URhGe}. Similarly,
the ferromagnetism in ZrZn$_2$ also shows a QCP at high pressures.
But in this case, unlike UGe$_2$, the highest
superconducting transition temperatures are observed at ambient
pressure,  that is at the furthest point from the
ferromagnetic-paramagnetic QCP.

Theoretically it is thought that at or near to the QCP quantum
spin fluctuations, can lead to spin-fluctuation induced pairing.
For the case of ferromagnetic QCP this was first studied by Fay
and Appel \cite{Fay&Appel} (who also suggested that ZrZn$_2$ might
be a suitable system in which to observe this effect). In this
case the ferromagnetic spin fluctuations lead to spin-triplet
pairing, by analogue with the case of superfluid $^3$He.  By
contrast, in the case of quantum critical antiferromagnetic spin
fluctuations spin-singlet d-wave pairing states are favoured
\cite{Bulut}.

Currently, very little is known about the superconducting
pairing state in the ferromagnetic superconductors UGe$_2$, URhGe and ZrZn$_2$.
If the pairing mechanism is indeed caused by ferromagnetic spin fluctuations, then
we might expect spin-triplet pairing states.  However, presently there
is insufficient evidence in support of this hypothesis to be decisive.
Thus it is still legitimate to consider other scenarios. In fact this is what we shall do here.
In short, we point out that the decline of the superconducting transition temperature, $T_c$,
with pressure could be a simple consequence of p-wave pairing of arbitrary origin in an exchange field.

In particular we will consider a simple model of the coexistence
between ferromagnetism and superconductivity based on a
parameterised electron-electron attractive interaction of
unspecified origin. We will derive Bogoliubov--de Gennes (BdG) and
gap equations for this model using the Hartree--Fock--Gorkov
approximation. We will consider separately the cases of: spin
singlet (s-wave) pairing, Opposite Spin Pairing (OSP) and Equal
Spin Pairing (ESP) spin triplet (p-wave) states.  Solving the gap
equations for these pairing states, we will then illustrate some
important properties of superconductivity in the presence of
ferromagnetism.

\section{A simple model for a ferromagnetic superconductor}

We  consider superconductivity arising in a Hubbard model with an
effective attractive pairwise interaction $U_{ij\sigma\sigma'}$,
acting between electrons at crystal sites $i$, $j$ with spins
$\sigma$ and $\sigma'$. In principle this effective interaction
could arise from either conventional pairing mechanisms, such as
electron-phonon coupling, or exchange of spin-fluctuations. Here
we shall assume that the effective interaction is both
short-ranged in space, namely $U_{ij\sigma\sigma'}\neq 0 $ only
for $i=j$ or nearest neighbours, and non-retarded.

In the ferromagnetic state we must also include the effective
exchange field caused by the ferromagnetism. This enters in the
model Hamiltonian as to the Zeeman splitting $\vec{V}_{xc}$. Thus
the complete Hamiltonian for this model is

\begin{equation}
\hat{\cal{H}}= -\sum_{ij\sigma}t_{ij}
% e^{-\I A_{ij}}
\hat{c}_{i\sigma}^\dag\hat{c}_{j\sigma}
+\frac{1}{2}\sum_{ij\sigma\sigma'}U_{ij\sigma\sigma'}\hat{n}_{i\sigma}\hat{n}_{j\sigma'}
+ \sum_{i\sigma\sigma'}\hat{c}_{i\sigma}^\dag
(\vec{\sigma}_{\sigma\sigma'}\cdot\vec{V}_{xc})\hat{c}_{i\sigma}
\end{equation}
where the $\hat{c}_{i\sigma}^{(\dag)}$ are the usual annihilation
(creation) operators, $\hat{n}_{i\sigma}$ is the number operator
and the $\vec{\sigma}_{\sigma\sigma'}$ are the components of the
vector of Pauli matrices

\begin{equation}
\underline{\underline{\vec{\sigma}}}=(\underline{\underline{\sigma}}_1,
\underline{\underline{\sigma}}_2,\underline{\underline{\sigma}}_3).
\end{equation}

In this context we should note that the ferromagnetism of ZrZn$_2$
is accurately described by the LSDA as a weak itinerant
ferromagnet. Experimentally the exchange splitting is clearly
resolved in de Haas-van Alphen experiments \cite{Yates} and band
structure calculations (also presented in reference \cite{Yates})
are in excellent agreement with these experiments. Moreover, the
calculated moment (0.18$\mu_B$) is close to the observed moment
(0.17$\mu_B$). Both the Curie temperature, $T_{FM}$, and low
temperature magnetisation are linear functions of pressure
\cite{Cordes}. Hence the low temperature magnetisation is a linear
function of $T_{FM}$, in line with the predictions of the Stoner
model. The most unusual magnetic property of \ZZ is that, although
a field of 0.05~T is enough to form a single magnetic domain, the
ordered moment is unsaturated up to 35~T
\cite{Pfleiderer,Deursen}. This is far more naturally understood
in an itinerant model such as LSDA or the Stoner model than, say,
the Heisenberg model. On the other hand, we hasten to add that it
is not clear whether this picture is useful for the ferromagnetic
superconductor UGe$_2$, since there the moments are much more
strongly localised.

Making the usual Hartree-Fock--Gorkov approximation, such that
$\Delta_{ij\sigma\sigma'} = -U_{ij\sigma\sigma'} \langle
\hat{c}_{i\sigma} \hat{c}_{j\sigma'} \rangle$, and using the
spin-generalised Bogoliubov--Valatin transformation,

\begin{equation}
\hat{c}_{i\sigma} = \sum_{\vec{k}\sigma'}
u_{\vec{k}\sigma\sigma'}(\vec{R}_i) \hat{\gamma}_{\vec{k}\sigma'}
+ v_{\vec{k}\sigma\sigma'}^*(\vec{R}_i)
\hat{\gamma}_{\vec{k}\sigma'}^\dagger
\label{eqn:Bogoliubov-Vlatin}
\end{equation}
subject to the completeness relation

\begin{equation}
\sum_{\vec{k}\sigma} \left( u_{\vec{k}\alpha\sigma}^*(\vec{R}_i)
u_{\vec{k}\beta\sigma}(\vec{R}_j) +
v_{\vec{k}\alpha\sigma}(\vec{R}_i)
v_{\vec{k}\beta\sigma}^*(\vec{R}_j) \right) =
\delta_{ij}\delta_{\alpha\beta}, \label{eqn:completeness}
\end{equation}
we find that the Bogoliubov de Gennes (BdG) equations
for this Hamiltonian are

\begin{eqnarray} \fl \left(
\begin{array}{cccc}
\varepsilon _{\vec k}+V_{xc3} & V_{xc1}-\I V_{xc2} & \Delta
_{\uparrow\uparrow }({\vec{k}}) & \Delta_{\uparrow\downarrow
}(\vec{k})
\\ V_{xc1}+\I V_{xc2} &\varepsilon _{\vec{k}}-V_{xc3} &
\Delta_{\downarrow\uparrow }({\vec{k}}) &
\Delta_{\downarrow\downarrow }({\vec{k}})
\\ -\Delta _{\uparrow\uparrow }^* (-\vec{k}) &
-\Delta_{\uparrow\downarrow }^* (-\vec{k}) &
-\varepsilon_{-\vec{k}}-V_{xc3} & -V_{xc1}-\I V_{xc2}
\\ -\Delta _{\downarrow\uparrow }^* (-\vec{k}) &
-\Delta_{\downarrow\downarrow }^* (-\vec{k}) & -V_{xc1}+\I V_{xc2}
& -\varepsilon_{-\vec{k}}+V_{xc3}
\end{array} \right) \left(
\begin{array}{c}
u_{\uparrow\sigma}({\vec{k}}) \\ u_{\downarrow\sigma}({\vec{k}}) \\
v_{\uparrow\sigma}({\vec{k}}) \\ v_{\downarrow\sigma}({\vec{k}})
\end{array} \right) \nonumber \\
= E_{\sigma}({\vec{k}}) \left(
\begin{array}{c}
u_{\uparrow\sigma}({\vec{k}}) \\ u_{\downarrow\sigma}({\vec{k}}) \\
v_{\uparrow\sigma}({\vec{k}}) \\ v_{\downarrow\sigma}({\vec{k}})
\end{array} \right) \label{eqn:spin gen BdG}
\end{eqnarray}
where $\varepsilon_{\vec{k}}$ is the normal (that is non
superconducting and non ferromagnetic) state energy and
$\vec{V}_{xc} = (V_{xc1},V_{xc2},V_{xc3})$.

The superconducting order parameter,
$\Delta_{\sigma\sigma'}(\vec{k})$, is calculated self-consistently
from

\begin{equation} \fl
\label{eqn:self_const} \Delta_{\sigma\sigma'}({\bf
k})=-\frac{1}{2}\sum_{\vec{q}\sigma''}U_{\sigma\sigma'}(\vec{k}-\vec{q})
\Big(u_{\sigma\sigma''}(-\vec{q})v^\ast_{\sigma'\sigma''}(-\vec{q})-
v^\ast_{\sigma\sigma''}(\vec{q})u_{\sigma'\sigma''}(\vec{q})\Big)(1-2f_{E_{\vec{q}\sigma''}}).
\end{equation}

We now introduce the Balain-Werthamer (BW) transformation
\cite{Balian&Werthamer,Vollhardt},

\begin{equation}
\underline{\underline{\Delta}}({\vec{k}})\equiv \left(
\begin{array}{cc}
\Delta
_{\uparrow\uparrow}(\vec{k})&\Delta_{\uparrow\downarrow}({\vec{k}})
\\ \Delta_{\downarrow\uparrow }({\vec{k}}) &\Delta_{\downarrow\downarrow }({\vec{k}})
\end{array} \right)
=\left(d_0(\vec{k})+\underline{\underline{\sigma}}\cdot\vec{d}(\vec{k})\right)\I\underline{\underline{\sigma_2}}.
\end{equation}
which separates the superconducting order parameter into a singlet
(scalar) part, $d_0(\vec{k})$ and a triplet (vector) part,
$\vec{d}(\vec{k})=(d_1(\vec{k}),d_2(\vec{k}),d_3(\vec{k}))$. In
terms of these parameters the BdG equations can be rewritten as

\begin{eqnarray} \fl \left(
\begin{array}{cccc}
\varepsilon _{\vec k}+V_{xc3} & V_{xc1}-\I V_{xc2} &
-d_{1}(\vec{k})+\I d_{2}({\vec{k}}) &
d_{0}({\vec{k}})+d_{3}({\vec{k}})
\\ V_{xc1}+\I V_{xc2} &\varepsilon _{\vec{k}}-V_{xc3} &
-d_{0}({\vec{k}})+d_{3}({\vec{k}}) & d_{1}({\vec{k}})+\I
d_{2}({\vec{k}})
\\ -d^\ast_{1}({\vec{k}})- \I d^\ast_{2}({\vec{k}}) & -d^\ast_{0}({\vec{k}})+d^\ast_{3}({\vec{k}}) &
-\varepsilon_{-\vec{k}}-V_{xc3} & -V_{xc1}-\I V_{xc2}
\\ d^\ast_{0}({\vec{k}})+d^\ast_{3}({\vec{k}}) & d^\ast_{1}({\vec{k}})- \I d^\ast_{2}({\vec{k}}) & -V_{xc1}+\I V_{xc2}
& -\varepsilon_{-\vec{k}}+V_{xc3}
\end{array} \right) \nonumber \\
\left(
\begin{array}{c}
u_{\uparrow\sigma}({\vec{k}}) \\ u_{\downarrow\sigma}({\vec{k}}) \\
v_{\uparrow\sigma}({\vec{k}}) \\ v_{\downarrow\sigma}({\vec{k}})
\end{array} \right)
= E_{\sigma}({\vec{k}}) \left(
\begin{array}{c}
u_{\uparrow\sigma}({\vec{k}}) \\ u_{\downarrow\sigma}({\vec{k}}) \\
v_{\uparrow\sigma}({\vec{k}}) \\ v_{\downarrow\sigma}({\vec{k}})
\end{array} \right)
\end{eqnarray}

Using this formalism, it is also possible to calculate the free energy in the general
case. This is given by

\begin{eqnarray} \fl
F = \sum_{\vec{k}\alpha\sigma} \varepsilon_\vec{k} \Big(
u_{\alpha\sigma}^*(-\vec{k}) u_{\alpha\sigma}(-\vec{k})
f_{\vec{k}\sigma} + v_{\alpha\sigma}(\vec{k})
v_{\alpha\sigma}^*(\vec{k}) \big(1-f_{\vec{k}\sigma}\big)
\nonumber \Big)
\\
 -\frac{1}{2} \sum_{\vec{k}\vec{k}'\alpha\beta\sigma\sigma'}
 \Bigg(
U(\vec{k}-\vec{k}') \nonumber \\ \times \Big[
u_{\alpha\sigma}^*(-\vec{k}) v_{\beta\sigma}(\vec{k})
f_{\vec{k}\sigma} + v_{\alpha\sigma}(-\vec{k})
u_{\beta\sigma}^*(\vec{k}) \big( 1
- f_{\vec{k}\sigma} \big) \Big]  \nonumber \\
  \times \Big[ u_{\alpha\sigma'}(-\vec{k}')
v_{\beta\sigma'}^*(\vec{k}') f_{\vec{k}'\sigma'} +
v_{\alpha\sigma'}^*(-\vec{k}') u_{\beta\sigma'}(\vec{k}') \big( 1
- f_{\vec{k}'\sigma'} \big) \Big] \Bigg) \nonumber \\
 + \sum_{\vec{k}\alpha\beta\sigma} \big(
\vec{\sigma}_{\alpha\beta} \cdot \vec{H} \big) \Big(
u_{\alpha\sigma}^*(\vec{k}) u_{\beta\sigma}(\vec{k})
f_{\vec{k}\sigma} + v_{\alpha\sigma}(\vec{k})
v_{\alpha\sigma}^*(\vec{k}) \big(1-f_{\vec{k}\sigma}\big) \Big)
\nonumber
\\  - k_BT \sum_{\vec{k}\sigma} \Big[ f_{\vec{k}\sigma}
\ln{f_{\vec{k}\sigma}} + \big(1 - f_{\vec{k}\sigma}\big) \ln{\big(
1-f_{\vec{k}\sigma} \big)} \Big] \label{eqn:free energy}
\end{eqnarray}

At this stage one must resort either to solving these equations
numerically \cite{Ben1,BenThesis}, or to studying special cases.
In this paper we shall take the later approach. First, we begin by
considering the case singlet pairing only (i.e. $d_1(\vec{k})=
d_2(\vec{k})= d_3(\vec{k}) =0$). In section \ref{section:triplet}
we will consider the case of only triplet pairing (i.e. when
$d_0(\vec{k}) =0$).

\section{The coexistence of singlet superconductivity and ferromagnetism}

In the case of a s-wave spin singlet superconductor, it was shown
by Fulde, Ferrel, Larkin and Ovchinnikov (FFLO)
\cite{Fulde&Ferrell,Larkin&Ovchinnikov} that the superconducting
ground state becomes non-uniform for small external exchange
fields. This solution is well known, and we shall not study it
here. On the other hand there are also solutions which are
spatially uniform. Whichever of these solutions is the ground
state can only be determined by calculating the free energy for
both and finding which is the lower solution. In strong fields the
FFLO state will be the minimum, but in weaker fields the FFLO
state will be unstable to  the uniform solution. In the rest of
this section we study the gap equations for the spatially uniform
case.

It is straightforward to show that $d_0(\vec{k})$ transforms as a
scalar under spin rotation. Thus, if there is no superconductivity in
the triplet channel, we can, without loss of generality, rewrite
the BdG equations as

\begin{eqnarray} \fl \left(
\begin{array}{cccc}
\varepsilon _{\vec k}+V_{xc} & 0 & 0 & d_{0}({\vec{k}})
\\ 0 &\varepsilon _{\vec{k}}-V_{xc} &
-d_{0}({\vec{k}}) & 0
\\ 0 & -d^\ast_{0}({\vec{k}}) &
-\varepsilon_{-\vec{k}}-V_{xc} & 0
\\ d^\ast_{0}({\vec{k}}) & 0 & 0
& -\varepsilon_{-\vec{k}}+V_{xc}
\end{array} \right)
\left(
\begin{array}{c}
u_{\uparrow\sigma}({\vec{k}}) \\ u_{\downarrow\sigma}({\vec{k}}) \\
v_{\uparrow\sigma}({\vec{k}}) \\ v_{\downarrow\sigma}({\vec{k}})
\end{array} \right) \nonumber \\
= E_{\sigma}({\vec{k}}) \left(
\begin{array}{c}
u_{\uparrow\sigma}({\vec{k}}) \\ u_{\downarrow\sigma}({\vec{k}}) \\
v_{\uparrow\sigma}({\vec{k}}) \\ v_{\downarrow\sigma}({\vec{k}})
\end{array} \right). \label{eqn:singlet BdG}
\end{eqnarray}
by rotating our spin reference frame so that
$V_{xc}=\sqrt{{V_{xc1}}^2+{V_{xc2}}^2+{V_{xc3}}^2}$.

\Eref{eqn:singlet BdG} can be separated into two sets of BdG
equations, so we have

\begin{equation}
\left(
\begin{array}{cc}
\varepsilon _{\vec{k}}+V_{xc} & d_{0}({\vec{k}})
\\ d^\ast_{0}({\vec{k}}) & -\varepsilon_{\vec{k}}+V_{xc}
\end{array} \right)
\left(
\begin{array}{c}
u_{\uparrow\uparrow}({\vec{k}}) \\
v_{\downarrow\uparrow}({\vec{k}})
\end{array} \right)
=E_{\uparrow}({\vec{k}}) \left(
\begin{array}{c}
u_{\uparrow\uparrow}({\vec{k}}) \\
v_{\downarrow\uparrow}({\vec{k}})
\end{array} \right) \label{eqn:singlet BdG a}
\end{equation}
and

\begin{equation}
\left(
\begin{array}{cc}
\varepsilon _{\vec{k}}-V_{xc} & -d_{0}({\vec{k}})
\\ -d^\ast_{0}({\vec{k}}) & -\varepsilon_{\vec{k}}-V_{xc}
\end{array} \right)
\left(
\begin{array}{c}
u_{\downarrow\downarrow}({\vec{k}}) \\
v_{\uparrow\downarrow}({\vec{k}})
\end{array} \right)
=E_{\downarrow}({\vec{k}}) \left(
\begin{array}{c}
u_{\downarrow\downarrow}({\vec{k}}) \\
v_{\uparrow\downarrow}({\vec{k}})
\end{array} \right). \label{eqn:singlet BdG b}
\end{equation}

\begin{figure}
    \centering
    \epsfig{figure=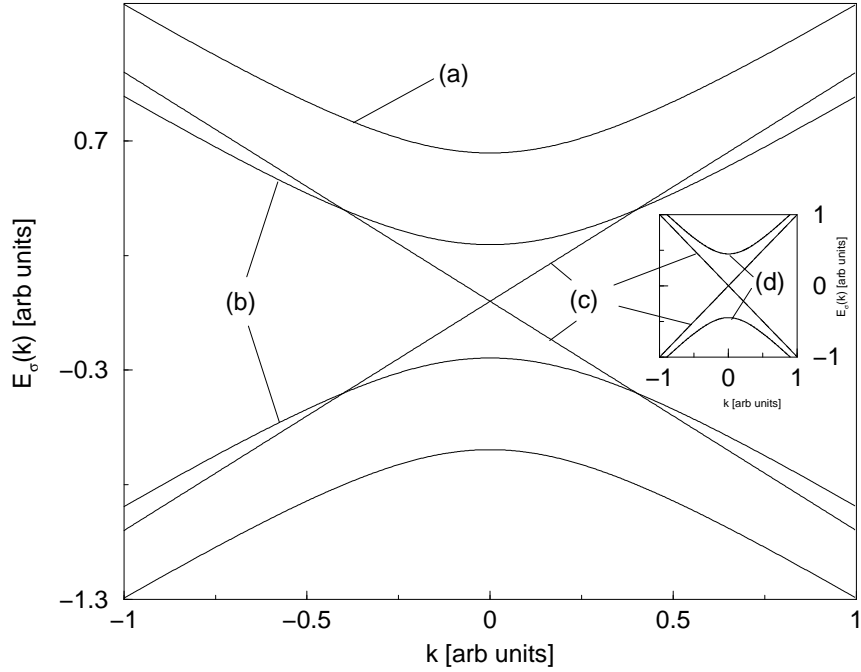, width=9cm, angle=270}
    \caption{The four branches of the singlet spectrum in a magnetic field. Inset,
    the zero field limit where the two spin branches become degenerate. The branches are (a) the spectra for
    $\sigma = \uparrow$, (b) the spectra for $\sigma = \downarrow$, (c) the normal state
    spectra in zero field and (d) the singlet spectrum for $V_{xc}=0$.} \label{fig:singlet spectrum}
\end{figure}

It is now a simple matter to regain the standard result
 \cite{Mineev&Samokhin} for the spectrum of a singlet
superconductor in a spin only magnetic field:

\begin{equation}
E_{\sigma}(\vec{k})=
\sqrt{\varepsilon_{\vec{k}}^2+|d_0(\vec{k})|^2} +
\sigma|\vec{V}_{xc}|, \label{eqn:singlet spectrum}
\end{equation}
with $\sigma = \uparrow\, \equiv 1$ and $\sigma = \downarrow\,
\equiv -1$.  The four corresponding energy levels are sketched in
Fig,~\ref{fig:singlet spectrum}.
\Eref{eqn:singlet spectrum} clearly reduces to the
standard BCS expression for the spectrum of a singlet
superconductor in the absence of exchange splitting as
$V_{xc}\rightarrow 0$. Also, when $V_{xc}=0$ equations
(\ref{eqn:singlet BdG a}) and (\ref{eqn:singlet BdG b}) reduce to
the usual BdG equations \cite{de_Gennes_SC} and we see that we are
justified in associating $d_0(\vec{k})$ with the usual singlet
superconducting order parameter $\Delta(\vec{k})$.

It is clear from \eref{eqn:singlet BdG} that

\begin{equation}
u_{\sigma-\sigma}(\vec{k}) = v_{\sigma\sigma}(\vec{k}) = 0,
\end{equation}
and it can also be shown that

\begin{equation}
u_{\sigma\sigma}(\vec{k}) =
\frac{d_0(\vec{k})}{\sqrt{(E_0(\vec{k})-\varepsilon_{\vec{k}})^2+|d_0(\vec{k})|^2}}
\end{equation}
and
\begin{equation}
v_{\sigma-\sigma}(\vec{k}) =
\frac{E_0(\vec{k})-\varepsilon_{\vec{k}}}{\sqrt{(E_0(\vec{k})-\varepsilon_{\vec{k}})^2+|d_0(\vec{k})|^2}}
\end{equation}
where
\begin{equation}
E_0(\vec{k})=\sqrt{\varepsilon_{\vec{k}}+|d_0(\vec{k})|^2}.
\end{equation}
$E_0(\vec{k})$ is, of course, of the same mathematical form as the
spectrum of a singlet superconductor in the absence of exchange
splitting. However, it is \emph{not} correct to say that
$E_0(\vec{k})$ \emph{is} the spectrum of a singlet superconductor
in the absence of exchange splitting as the value of
$d_0(\vec{k})$ (although, importantly, not the value of
$\varepsilon(\vec{k})$) depends on $V_{xc}$ in general.

Substituting our expressions for the eigenvectors of the BdG into
the self-consistency condition (\ref{eqn:self_const}) we find that
the gap equation is

\begin{equation}
d_0(\vec{k})=-\frac{1}{4}\sum_{\vec{k}\sigma}U_{\sigma-\sigma}(\vec{k})\frac{d_0(\vec{k})}
{E_0(\vec{k})}\tanh\left(\frac{E_0(\vec{k})+\sigma
V_{xc}}{2k_BT}\right). \label{eqn:singlet gap eqn}
\end{equation}
In the absence of exchange splitting the gap equation regains its
familiar BCS form \cite{Ketterson&Song}. However, we note that
surprisingly the exchange splitting dependence of the gap only
enters via the Fermi ($\tanh$) term. This means that when $T=0$
the gap equation becomes

\begin{equation}
d_0(\vec{k})=-\frac{1}{4}\sum_{\vec{k}\sigma}U_{\sigma-\sigma}(\vec{k})\frac{d_0(\vec{k})}
{E_0(\vec{k})}.
\end{equation}
which is independent of $\vec{V}_{xc}$.

We must now ask what this result means physically.
%, as, thus far,
%we have considered the gap equation purely as a mathematical exercise.
The most obvious conclusion is that, at zero
temperature, the gap in independent of exchange splitting. This is
true, but with one condition, which we will discuss below.

The gap equation is a non-linear integral equation. And, as such,
has, in general, more than one solution. (For example the trivial
solution $d_0(\vec{k}) = 0$ is always a solution.) All that we
have actually shown is that for any given solution $d_0(\vec{k})$
is independent $\vec{V}_{xc}$ at $T=0$. To find the ground state
we must consider all possible solutions and calculate the free
energy of each solution. In the absence of exchange splitting the
gap equation can be derived by minimising the free energy with
respect to the superconducting order parameter
\cite{GyorffyStauton&Stocks}. This leads to the conclusion that
the trivial solution is only the ground state when no other
solution exists. However, no such proof exists for a
superconductor in a finite exchange splitting. This means that it
is perfectly possible there to be a phase transition from the
superconducting to normal states as the exchange splitting is
increased at zero temperature. Any such phase transition will be
\lq perfectly' first order in the sense that the order parameter
will jump from zero (above the critical exchange splitting,
$V_{xc}^C$) to some finite value (below $V_{xc}^C$) and remain at
that value for all $V_{xc} \le V_{xc}^C$. The order
parameter as a function of exchange splitting will therefore
resemble a Heaviside step function. Of course, as in general other
superconducting phases can exist (such as the FFLO state) phase transitions can also occur
between different superconducting phases in a similar manner
\cite{Ben4}.

Such a phase transition was first studied independently by
Clogston \cite{Clogston} and Chandrasekhar \cite{Chandrasekhar}
who both, in fact, assumed the independence of $d_0(\vec{k})$ on
$V_{xc}$ that we have derived above. Using this assumption they
were able to show from simple thermodynamics that if the exchange
splitting is greater than $V_{xc}^P \equiv |\Delta(0)|/\sqrt{2}$
where $|\Delta(0)|$ is the superconducting gap at zero temperature
(and zero exchange splitting) then the normal state has a lower
energy than the s-wave superconducting state. This is known both
of as Clogston--Chandrasekhar limiting and as Pauli-paramagnetic
limiting. Clogston--Chandrasekhar limiting clearly applies to all
singlet states, but does not necessarily apply to triplet states.
In most superconducting materials $\mu_BH_{C2} < V_{xc}^P$.
Therefore, if a superconductor has a large upper critical field in
comparison to the Clogston--Chandrasekhar limit this is good
evidence for triplet superconductivity. The FFLO state can also
display $\mu_BH_{C2} > V_{xc}^P$. Clogston--Chandrasekhar limiting
has been observed in the layered organic compound
$\kappa-$(BEDT-TTF)$_2$Cu(SCN)$_2$ \cite{Zuo} when a magnetic
field is applied parallel to the layers (which prevents the
formation of orbital currents due to the highly two dimensional
nature of the material).

\begin{figure}
    \centering
    \epsfig{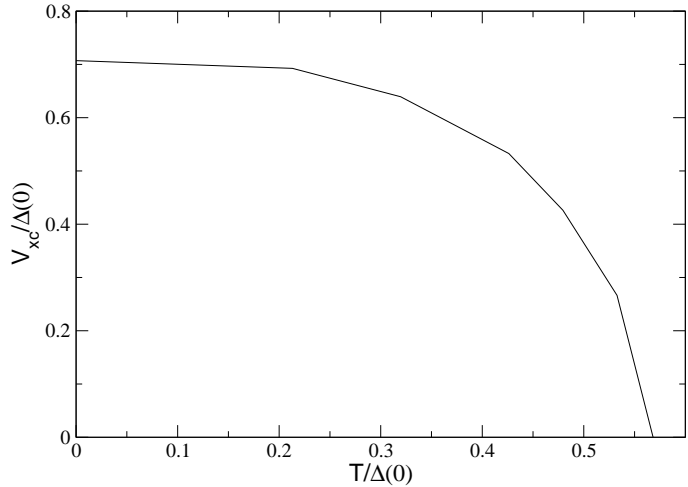}
    \caption{The phase diagram of an s-wave superconductor in an exchange  field
    calculated by
    solving the spin generalised BdG equations self consistently. Note that as the
    phase transition is first order in the presence of exchange splitting
    the free energy must be calculated for both the normal and superconducting
    states to correctly construct this phase diagram.}
\label{figure:s wave phase diagram}
\end{figure}

To illustrate this point, we have solved
 the gap equation \eref{eqn:singlet gap eqn} numerically
 for a cubic lattice.  We assumed
$U_{ij\sigma\sigma}=U\delta_{ij}$ (i.e. an on-site interaction)
corresponding to the case of local s-wave pairing. The comparison
between the calculated superconducting and normal state free
energies, leads to the phase diagram given in \fref{figure:s wave
phase diagram}. This  calculated phase diagram is in excellent
agreement with that calculated from the Baltensperger--Sarma
equation \cite{Baltensperger,Sarma}. However, while the
Baltensperger--Sarma equation only allows for the calculation of
the superconducting-metal phase transition, our numerical gap
equation solution allows for the evaluation of the order parameter
at any point in $T-V_{xc}$ space and hence for the evaluation of
thermodynamic variables such as the  heat capacity,

\begin{equation}
C_V = \sum_{\vec{k}\sigma} \frac{1}{k_BT^2} f_{\vec{k}\sigma}
\big( 1-f_{\vec{k}\sigma} \big) \left( E_{\vec{k}\sigma}^2 -
\frac{1}{E_{\vec{k}\sigma}-\sigma V_{xc}}\frac{\partial
|d_0(\vec{k})|^2}{\partial T} \right).
\end{equation}
and the magnetisation \cite{Mineev&Samokhin},

\begin{eqnarray}
M = -\frac{V}{(2\pi)^3} \sum_{\sigma}\sigma \int \D^3 \vec{k}
\frac{1}{1+e^{(E_{\vec{k}\sigma} - \mu)/k_BT} }.
\end{eqnarray}
where $V$ is the volume of the first Brillouin zone.

A numerical study of these equations \cite{BenThesis} shows that,
in an exchange field, the thermodynamic functions \lq see' an
effective gap, $\Delta_{eff}$, i.e.

\begin{equation}
\{C_V,M,\chi\}\sim e^{-\frac{\Delta_{eff}}{k_BT}}
\end{equation}
where

\begin{equation}
\Delta_{eff} = |\Delta(0)| - |\vec{V}_{xc}|.
\end{equation}

\section{The coexistence of triplet superconductivity and
ferromagnetism}\label{section:triplet}

We will now consider the properties of a triplet superconductor in
a magnetic field. Using a similar approach to the singlet
case above we  are able to derive many of the same physical quantities.
This highlights both similarities and differences between the singlet and
triplet cases, which may perhaps help in identifying the pairing state symmetry
in specific ferromagnetic superconductors.

Before we begin we will generalise a useful theorem due to de
Gennes \cite{de_Gennes_SC}. We begin by writing the BdG equations
(\ref{eqn:spin gen BdG}) in a pseudo-spinor notation:

\begin{equation}
\left(
\begin{array}{cc}
\underline{\underline{\xi}}({\vec{k}}) &
\underline{\underline{\Delta}}_{\vec{k}}
\\ -\underline{\underline{\Delta}}^*_{-\vec{k}} &
-\underline{\underline{\xi}}^*({\vec{k}})
\end{array}\right)
\left(
\begin{array}{c}
\underline{u}_\sigma({\vec{k}}) \\
\underline{v}_\sigma({\vec{k}})
\end{array}\right)
=E_\sigma({\vec{k}}) \left(
\begin{array}{c}
\underline{u}_\sigma({\vec{k}}) \\
\underline{v}_\sigma({\vec{k}})
\end{array}\right). \label{eqn:spinor BdG}
\end{equation}
Where

\begin{eqnarray}
\underline{\underline{\xi}}({\vec{k}}) &=& \left(
\begin{array}{cc}
\varepsilon _{\vec k}+V_{xc3} & V_{xc1}-\I V_{xc2} \\
V_{xc1}+\I V_{xc2} & \varepsilon _{\vec{k}}-V_{xc3}
\end{array}\right), \\
\underline{\underline{\Delta}}_{\vec{k}} &=& \left(
\begin{array}{cc}
\Delta _{\uparrow\uparrow }({\vec{k}}) &
\Delta_{\uparrow\downarrow}(\vec{k}) \\
\Delta_{\downarrow\uparrow }({\vec{k}}) &
\Delta_{\downarrow\downarrow }({\vec{k}})
\end{array}\right), \\
\underline{u}_\sigma({\vec{k}}) &=& \left(
\begin{array}{cc}
u_{\uparrow\sigma}({\vec{k}}) \\ u_{\downarrow\sigma}({\vec{k}})
\end{array}\right),
\end{eqnarray}
and
\begin{eqnarray}
\underline{v}_\sigma({\vec{k}}) &=& \left(
\begin{array}{cc}
v_{\uparrow\sigma}({\vec{k}}) \\ v_{\downarrow\sigma}({\vec{k}})
\end{array}\right).
\end{eqnarray}
Multiplying by $-1$, taking the complex conjugate, parity
inverting and exchanging the rows of \eref{eqn:spinor BdG} leads
to

\begin{equation}
\left( \begin{array}{cc}
\underline{\underline{\xi}}({\vec{k}}) &
\underline{\underline{\Delta}}({\vec{k}})
\\ -\underline{\underline{\Delta}}^*({\vec{k}}) &
-\underline{\underline{\xi}}^*({\vec{-k}})
\end{array}\right)
\left( \begin{array}{c}
\underline{u}^*_\sigma({\vec{-k}}) \\
\underline{v}^*_\sigma({\vec{-k}})
\end{array}\right)
= -E_\sigma({\vec{k}}) \left( \begin{array}{c}
\underline{u}^*_\sigma({\vec{-k}}) \\
\underline{v}^*_\sigma({\vec{-k}})
\end{array}\right),
\end{equation}
as both $E_\sigma({\vec{k}})$ and
$\underline{\underline{\xi}}({\vec{k}})$ are even under parity
inversion.

We have therefore shown that if $\left( \begin{array}{c} \underline{u}_\sigma({\vec{k}}) \\
\underline{v}_\sigma({\vec{k}}) \end{array}\right)$ is an
eigenvector of the spin-generalised BdG equations in a magnetic
field, with the corresponding eigenvalue $E_\sigma({\vec{k}})$
then, $\left( \begin{array}{c} \underline{u}^*_\sigma({\vec{-k}})
\\ \underline{v}^*_\sigma({\vec{-k}}) \end{array}\right)$ is also an
eigenvector and that the corresponding eigenvalue is
$-E_\sigma({\vec{k}})$. As $\sigma$ can take two values
($\uparrow$ or $\downarrow$) we have identified all of the
eigenstates.

This analysis holds for both triplet and singlet states. (For a
singlet state with $|\vec{V}_{xc}|=0$ it clearly reduces to the
theorem of de Gennes.) The spectrum for a singlet superconductor
in an exchange field (shown in \fref{fig:singlet spectrum}) is
clearly in agreement with this theorem.

When studying triplet states, and particularly when studying the
effect of exchange splitting on the triplet state, it is useful to
introduce the notion of unitary and non-unitary states. For a
triplet state

\begin{equation}
\underline{\underline{\Delta}}_\vec{k} \,
\underline{\underline{\Delta}}_\vec{k}^\dagger =
\underline{\underline{I}} \, |\vec{d}(\vec{k})|^2 + \I
\vec{\underline{\underline{\sigma}}} \cdot (\vec{d}(\vec{k})
\times \vec{d}(\vec{k})^*) \label{eqn:nonunitary gap}
\end{equation}
and, in the absence of exchange splitting,

\begin{equation}
E_\sigma(\vec{k}) = \sqrt{\varepsilon_\vec{k}^2 +
|\vec{d}(\vec{k})|^2 + \sigma |\vec{d}(\vec{k}) \times
\vec{d}(\vec{k})^*|}. \label{eqn:trip spec zer field}
\end{equation}

It is therefore useful to introduce the vector $\vec{q}(\vec{k})$
which is defined by

\begin{equation}
\vec{q}(\vec{k}) = \I \vec{d}(\vec{k}) \times \vec{d}(\vec{k})^*.
\end{equation}
It is clear that $\vec{q}(\vec{k})$ is a \emph{real} vector. A
unitary state is defined as any state in which
$\vec{q}(\vec{k})=0$ for all $\vec{k}$.

By setting the singlet order parameter, $d_0(\vec{k})$, to zero we
can write down the BdG equations for a triplet superconductor in
an exchange field,

\begin{eqnarray} \fl
\left(\begin{array}{cccc} \varepsilon _{\vec{k}}+V_{xc3} &
V_{xc1}-\I V_{xc2} & -d_{1}(\vec{k})+\I d_{2}({\vec{k}}) &
d_{3}({\vec{k}})
\\ V_{xc1}+\I V_{xc2} & \varepsilon _{\vec{k}}-V_{xc3} &
d_{3}({\vec{k}}) & d_{1}({\vec{k}})+\I d_{2}({\vec{k}})
\\ -d^\ast_{1}({\vec{k}})-\I d^\ast_{2}({\vec{k}}) &
d^\ast_{3}({\vec{k}}) &-\varepsilon_{\vec{k}}-V_{xc3} &
-V_{xc1}-\I V_{xc2}
\\ d^\ast_{3}({\vec{k}}) & d^\ast_{1}({\vec{k}})- \I d^\ast_{2}({\vec{k}})
& -V_{xc1}+\I V_{xc2}&-\varepsilon_{\vec{k}}+ V_{xc3}
\end{array}\right) \nonumber
\\ \left(\begin{array}{c}
u_{\uparrow\sigma}({\vec{k}}) \\ u_{\downarrow\sigma}({\vec{k}}) \\
v_{\uparrow\sigma}({\vec{k}}) \\ v_{\downarrow\sigma}({\vec{k}})
\end{array}\right)  =E_{\sigma}({\vec{k}})
\left(\begin{array}{c}
u_{\uparrow\sigma}({\vec{k}}) \\ u_{\downarrow\sigma}({\vec{k}}) \\
v_{\uparrow\sigma}({\vec{k}}) \\ v_{\downarrow\sigma}({\vec{k}})
\end{array}\right). \label{eqn:triplet BdG} \hspace*{15pt}
\end{eqnarray}

The eigenvalues of these BdG equations are given by \cite{BenThesis}

\begin{equation}
E_\sigma(\vec{k}) = \sqrt{\varepsilon_{\vec{k}}^2 +
\mu_B^2|\vec{V}_{xc}|^2 + |\vec{d}(\vec{k})|^2 +
\sigma\sqrt{\Lambda(\vec{k})}} \label{spectrum}
\end{equation}
where

\begin{equation}
\Lambda(\vec{k}) = |\vec{q}(\vec{k})|^2 +
4\varepsilon_{\vec{k}}^2\mu_B^2|\vec{V}_{xc}|^2 +
4\mu_B^2|\vec{V}_{xc}\cdot\vec{d}(\vec{k})|^2  +
4\varepsilon_{\vec{k}}\mu_B\vec{V}_{xc}\cdot\vec{q}(\vec{k}).
\end{equation}
In zero field we clearly have the usual result \eref{eqn:trip spec
zer field} for the spectrum a triplet superconductor.

Again, we can also derive the expressions for thermodynamic quantities in
a general triplet state. For example the heat capacity is given by

\begin{eqnarray}
C_V =
\sum_{\vec{k}\sigma}\frac{f_{\vec{k}\sigma}(1-f_{\vec{k}\sigma})}{k_BT^2}
\left(E_\sigma(\vec{k})^2-\frac{T}{2}\frac{d}{dT}|\vec{d}(\vec{k})|^2\right)
\end{eqnarray}
and the (vector) magnetisation, $\vec{M}$, is given by

\begin{eqnarray} \fl
\vec{M} = \sum_{\vec{k}} \bigg( u_{\uparrow\sigma}^*(\vec{k})
u_{\downarrow\sigma}(\vec{k}) f_{\vec{k}\sigma} +
v_{\uparrow\sigma}(\vec{k}) v_{\downarrow\sigma}^*(\vec{k})
\big(1-f_{\vec{k}\sigma}\big) \nonumber \\
+ u_{\downarrow\sigma}^*(\vec{k}) u_{\uparrow\sigma}(\vec{k})
f_{\vec{k}\sigma} + v_{\downarrow\sigma}(\vec{k})
v_{\uparrow\sigma}^*(\vec{k})
\big(1-f_{\vec{k}\sigma}\big),  \nonumber \\
-\I u_{\uparrow\sigma}^*(\vec{k}) u_{\downarrow\sigma}(\vec{k})
f_{\vec{k}\sigma} - \I v_{\uparrow\sigma}(\vec{k})
v_{\downarrow\sigma}^*(\vec{k})
\big(1-f_{\vec{k}\sigma}\big)  \nonumber \\
+ \I u_{\downarrow\sigma}^*(\vec{k}) u_{\uparrow\sigma}(\vec{k})
f_{\vec{k}\sigma} + \I v_{\downarrow\sigma}(\vec{k})
v_{\uparrow\sigma}^*(\vec{k}) \big(1-f_{\vec{k}\sigma}\big)
\nonumber, \\
+ u_{\uparrow\sigma}^*(\vec{k}) u_{\uparrow\sigma}(\vec{k})
f_{\vec{k}\sigma} + v_{\downarrow\sigma}(\vec{k})
v_{\downarrow\sigma}^*(\vec{k})
\big(1-f_{\vec{k}\sigma}\big) \nonumber \\
- u_{\downarrow\sigma}^*(\vec{k}) u_{\uparrow\sigma}(\vec{k})
f_{\vec{k}\sigma} - v_{\downarrow\sigma}(\vec{k})
v_{\uparrow\sigma}^*(\vec{k}) \big(1-f_{\vec{k}\sigma}\big)
\bigg). \label{eqn:magnetisation}
\end{eqnarray}

Following the methods of Sigrist and Ueda \cite{Sigrist&Ueda} it
can be shown \cite{BenThesis} that in the absence of exchange
splitting the gap equations for a triplet superconductor are

\begin{eqnarray} \fl
\Delta_{\alpha\beta}(\vec{k}) = \sum_{\vec{k}'}
U_{\alpha\beta}(\vec{k}-\vec{k}') \left[ \frac{1}{4
E_{\vec{k}\uparrow}} \left( \vec{d}(\vec{k}) + \I
\frac{\vec{q}(\vec{k}) \times
\vec{d}(\vec{k})}{|\vec{q}(\vec{k})|} \tanh \left( \frac{\beta
E_{\vec{k}\uparrow}}{2} \right)
 \right)  \right. \nonumber \\
 \left. + \frac{1}{4
E_{\vec{k}\downarrow}} \left( \vec{d}(\vec{k}) - \I
\frac{\vec{q}(\vec{k}) \times
\vec{d}(\vec{k})}{|\vec{q}(\vec{k})|} \tanh \left( \frac{\beta
E_{\vec{k}\downarrow}}{2} \right)
 \right) \right]. \label{eqn:gap eqn non unitary no exchange
 splitting}
\end{eqnarray}
However, these methods do not generalise to a finite exchange
splitting. Fortunately triplet states can be separated into three
classes: those that contain only OSP states, those that contain
only ESP states and those that contain both OSP and ESP states.
The first two cases represent a great simplification and we will
now study these special cases. However, it should be noted that
neither of the formalisms presented below can deal with states
that contain both OSP and ESP such as the B and B$_2$ phases.

\subsection{Opposite spin pairing}

An OSP state is defined as any state for which $\vec{d}(\vec{k})
\times \vec{V}_{xc} = 0$ for all $\vec{k}$. Thus, in this limited
sense, we may describe $\vec{d}(\vec{k})$ as parallel to
$\vec{V}_{xc}$. Much as in the case of singlet pairing we can,
without loss of generality, rotate the system, recalling the
$\vec{d}(\vec{k})$ transforms as a vector under rotation, so that

\begin{eqnarray} \fl \left(
\begin{array}{cccc}
\varepsilon _{\vec k}+V_{xc} & 0 & 0 & d_{3}({\vec{k}})
\\ 0 &\varepsilon _{\vec{k}}-V_{xc} &
d_{3}({\vec{k}}) & 0
\\ 0 & d^\ast_{3}({\vec{k}}) &
-\varepsilon_{-\vec{k}}-V_{xc} & 0
\\ d^\ast_{3}({\vec{k}}) & 0 & 0
& -\varepsilon_{-\vec{k}}+V_{xc}
\end{array} \right)
\left(
\begin{array}{c}
u_{\uparrow\sigma}({\vec{k}}) \\ u_{\downarrow\sigma}({\vec{k}}) \\
v_{\uparrow\sigma}({\vec{k}}) \\ v_{\downarrow\sigma}({\vec{k}})
\end{array} \right) \nonumber \\
= E_{\sigma}({\vec{k}}) \left(
\begin{array}{c}
u_{\uparrow\sigma}({\vec{k}}) \\ u_{\downarrow\sigma}({\vec{k}}) \\
v_{\uparrow\sigma}({\vec{k}}) \\ v_{\downarrow\sigma}({\vec{k}})
\end{array} \right). \label{eqn:para BdG}
\end{eqnarray}

Again we can separate \eref{eqn:para BdG} into two BdG equations
and hence, in a similar manner to which we derived the singlet gap
equation, we find that the gap equations for OSP triplet
superconductivity are

\begin{equation}
d_3(\vec{k})=-\frac{1}{4}\sum_{\vec{k}\sigma}U_{\sigma-\sigma}(\vec{k})\frac{d_3(\vec{k})}
{E_0(\vec{k})}\tanh\left( \frac{E_0(\vec{k})+\sigma V_{xc}
}{2k_BT}\right).
\end{equation}
Note that this equation is of precisely the same mathematical form
as the singlet gap equation \eref{eqn:singlet gap eqn}. Both the
phase diagram and the effective gap \lq seen' by thermodynamic
probes are the same as we earlier found for singlet
superconductivity. However, this time the effective gap \lq seen'
by thermodynamic probes is given by \cite{Ben1,BenThesis}

\begin{equation}
\Delta_{eff}=\overline{|\vec{d}(\vec{k_F})|}-|\vec{V}_{xc}|.
\end{equation}
where $\overline{|\vec{d}(\vec{k_F})|}$ is the mean gap at the
Fermi surface.

All singlet pairing states are, by definition, OSP states. Thus it
appears that, in the presence of exchange splitting, the important
property of a state is whether it is an OSP or an ESP state, not
whether it is a triplet or a singlet state.

\subsection{Equal spin pairing}

An ESP state is defined as any state for which $\vec{d}(\vec{k})
\cdot \vec{V}_{xc} = 0$ for all $\vec{k}$. Thus, in this limited
sense, we may describe $\vec{d}(\vec{k})$ as perpendicular to
$\vec{V}_{xc}$. In this case, for $\vec{V}_{xc} = \big(
0,0,-V_{xc} \big)$, the spin triplet BdG equations are

\begin{eqnarray} \fl \left(
\begin{array}{cccc}
\varepsilon _{\vec k}-V_{xc} & 0 &
\Delta_{\uparrow\uparrow}({\vec{k}}) & 0
\\ 0 & \varepsilon _{\vec{k}}+V_{xc} & 0
& \Delta_{\downarrow\downarrow }({\vec{k}})
\\ -\Delta_{\uparrow\uparrow }^*(-\vec{k})
& 0 & -\varepsilon_{-\vec{k}}+V_{xc} & 0
\\ 0 & -\Delta_{\downarrow\downarrow}^*(-\vec{k})
& 0 & -\varepsilon_{-\vec{k}}-V_{xc}
\end{array} \right)
\left(
\begin{array}{c}
u_{\uparrow\sigma}({\vec{k}}) \\ u_{\downarrow\sigma}({\vec{k}}) \\
v_{\uparrow\sigma}({\vec{k}}) \\ v_{\downarrow\sigma}({\vec{k}})
\end{array} \right) \nonumber \\
= E_{\sigma}({\vec{k}}) \left(
\begin{array}{c}
u_{\uparrow\sigma}({\vec{k}}) \\ u_{\downarrow\sigma}({\vec{k}}) \\
v_{\uparrow\sigma}({\vec{k}}) \\ v_{\downarrow\sigma}({\vec{k}})
\end{array} \right).
\end{eqnarray}

We can now easily separate the BdG equations into a pair of BdG
equations for up electrons,

\begin{eqnarray}\left(
\begin{array}{cc}
\varepsilon _{\vec k}-V_{xc} &
\Delta_{\uparrow\uparrow}({\vec{k}})
\\ -\Delta_{\uparrow\uparrow }^*(-\vec{k})
& -\varepsilon_{-\vec{k}}+V_{xc}
\end{array} \right) \left(
\begin{array}{c}
u_{\uparrow\sigma}({\vec{k}}) \\ v_{\uparrow\sigma}({\vec{k}})
\end{array} \right)
= E_{\sigma}({\vec{k}}) \left(
\begin{array}{c}
u_{\uparrow\sigma}({\vec{k}}) \\ v_{\uparrow\sigma}({\vec{k}}) \\
\end{array} \right).
\end{eqnarray}
and a set of BdG equations for down electrons,

\begin{eqnarray} \left(
\begin{array}{cc}
\varepsilon _{\vec{k}}+V_{xc} & \Delta_{\downarrow\downarrow
}({\vec{k}})
\\ -\Delta_{\downarrow\downarrow}^*(-\vec{k})
& -\varepsilon_{-\vec{k}}-V_{xc}
\end{array} \right)
\left(
\begin{array}{c}
u_{\downarrow\sigma}({\vec{k}}) \\
v_{\downarrow\sigma}({\vec{k}})
\end{array} \right)
= E_{\sigma}({\vec{k}}) \left(
\begin{array}{c}
u_{\downarrow\sigma}({\vec{k}}) \\
\end{array} \right).
\end{eqnarray}
Using the self-consistency condition \eref{eqn:self_const} we
easily find that the gap equations are

\begin{eqnarray}
\Delta_{\sigma\sigma}({\vec{k}}) = -\sum_{\vec{k}'}
\frac{U_{\sigma\sigma}(\vec{k}-\vec{k}')
\Delta_{\sigma\sigma}({\vec{k}'})}{2 E_\sigma(\vec{k}')}
(1-2f_{E_{\vec{k}'\sigma}}). \label{eqn:ESP gap eqn nice form}
\end{eqnarray}
with

\begin{eqnarray}
E_{\vec{k}\sigma} = \sqrt{\left( \varepsilon_{\vec{k}} - \sigma
V_{xc} \right) ^2 + |\Delta_{\sigma\sigma}({\vec{k}})|^2}
\end{eqnarray}

As $T \rightarrow T_C$ from below,
$|\underline{\underline{\Delta}}_\vec{k}| \rightarrow 0$ and hence
$E_\sigma (\vec{k}) \rightarrow \varepsilon (\vec{k}) + \sigma
V_{xc}$. Therefore the gap equation becomes

\begin{eqnarray}
\Delta_{\sigma\sigma}({\vec{k}}) = \sum_{\vec{k}'}
\frac{U_{\sigma\sigma}(\vec{k}-\vec{k}') }{2 \big( \varepsilon
(\vec{k}') - \sigma V_{xc} \big)} \tanh \left( \frac{\varepsilon
(\vec{k}') - \sigma V_{xc}}{2k_BT} \right)
\Delta_{\sigma\sigma}({\vec{k}'}). \label{eqn:linearised gap eqn}
\end{eqnarray}
Thus, near $T_C$ the gap equation is linear. This allows $T_C$ to
be determined very accurately. Further by comparing the transition
temperatures of various symmetries one can find which has the
highest transition temperature and hence which state occurs for $T
\lesssim T_C$.

Clearly, one cannot, in general, use the linearised gap equation
to study transitions from one superconducting state to another as
the gap equation can no longer be linearised below the first
superconducting transition. The exception to this rule is the
transition from an ESP state with only one type of pairing to an
ESP state with both $\uparrow\uparrow$ and $\downarrow\downarrow$
pairing (an example of such a transition is the transition from
the A$_1$ phase to the A$_2$ phase), because of the complete
separation of the spin-up and spin-down subsystems in the presence
of exchange splitting and the absence of opposite spin pairing or
spin flip processes.

We solved the linearised gap equations (\ref{eqn:linearised gap
eqn}) numerically for parameters chosen of \zz (see \cite{Ben2}
for a discussion). To do this we used a simple cubic tight binding
model and a k-space integration mesh of $10^9$ points. We use such
a large array for two reasons. A fine integration mesh is required
to accurately determine the density of states (DOS). Our method
(implicitly) requires an accurate calculation of the spin
dependant DOS, $D_\sigma(\varepsilon_F)$. This is particularly
important in our case as we are varying the exchange splitting and
thus we are changing the $D_\sigma(\varepsilon_F)$, so any errors
in evaluating $D_\sigma(\varepsilon)$ will lead to significant
errors in our calculation of the variation of $T_C$ with $V_{xc}$.

\begin{figure}
    \centering
    \epsfig{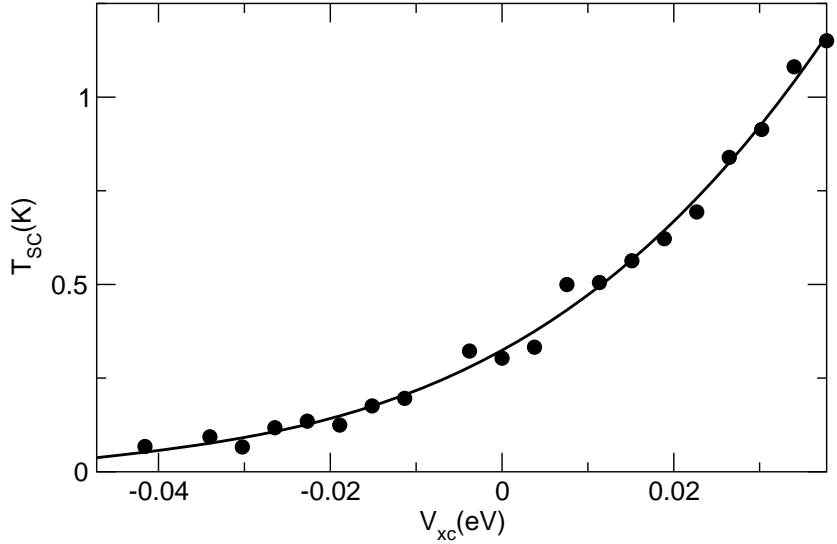}
    \caption{The results of our numerical solution of the linearised
    gap equations are shown by the points. The line is a fit to the
    calculated points by a cubic equation.} \label{fig:numerical_res_zz_fit}
\end{figure}

We show the results of our numerical calculations in figure
\ref{fig:numerical_res_zz_fit}. The line is a cubic curve fitted
to the numerical data. For any given exchange splitting, $V_{xc}$,
there are two transition temperatures, corresponding to the two
separate spin components of the ESP order parameter. We have
plotted the transition temperature for $\uparrow\uparrow$ pairing
on the positive $V_{xc}$ side of the graph and the transition
temperature for $\downarrow\downarrow$ paring on the negative
$V_{xc}$ scale. There are several reasons for plotting the data in
this way.

\vspace*{0.2cm}(i) In this way the graph shows the behaviour of the $\uparrow\uparrow$ pairing
state over a full range of exchange splitting, from positive to negative.

\vspace*{0.2cm}(ii) We see that the point $V_{xc}=0$  is not a
special case, and the curve is smooth there.

\vspace*{0.2cm}(iii) We also have a
larger data range to fit over,  and thus increase the accuracy of the
cubic fit.

\vspace*{0.2cm} Zero exchange splitting is not a special point
because in both the non-linear and linearised gap equations
exchange splitting is mathematically equivalent to a change in
chemical potential. Thus, the graph   plotted in the manner shown
in Fig. \ref{fig:numerical_res_zz_fit} can also be interpreted as
a plot of critical temperature of the of the triplet A phase as a
function of the chemical potential in zero exchange splitting.

\begin{figure}
    \centering
    \epsfig{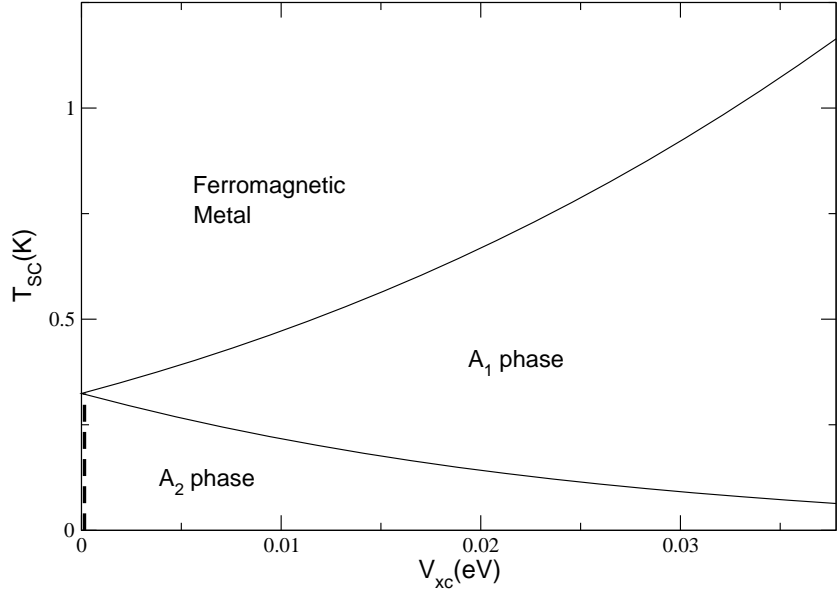}
    \caption{The phase diagram of our model. The critical temperature is shown for both
    A$_1$ and A$_2$ phases over a range of exchange splittings.
    The hatched area indicates the A phase, which is the ground state when $V_{xc}=0$.}
    \label{fig:numerical_res_zz}
\end{figure}

We now plot the critical temperature for both $\uparrow\uparrow$
and $\downarrow\downarrow$ pairing on the same graph (figure
\ref{fig:numerical_res_zz}). This plot shows is then the
($V_{xc}$, $T$)  superconducting phase diagram for our model.
(This, of course, assumes that no further phase transitions occur
at low temperatures.) The higher transition temperature is the
transition to the A$_1$ phase (where only $\uparrow \uparrow$ pairing occurs)
and the second transition is a
transition to the A$_2$ phase (where $\downarrow \downarrow$ pairing begins).
In the paramagnetic state (the line
$V_{xc}=0$) the superconducting state is an A phase as the
superconducting order parameter is the same for both the
$\uparrow\uparrow$ and $\downarrow\downarrow$ pairing states. (The
A$_2$ phase becomes the A phase via a cross over, rather than a
phase transition.)

The phase diagram shown in figure \ref{fig:numerical_res_zz} is
clearly equivalent to the A$_1$-A$_2$ splitting of \He in a
magnetic field. Experimental measurement of this phase transition
in \He due to Remeijer \etal are reported reference
\cite{Remeijer}. At first sight figure \ref{fig:numerical_res_zz}
and reference \cite{Remeijer} appear rather different, however the
are in fact almost identical, as we will now show. The
dimensionless measure of the exchange splitting for the Remijer
\etal experiments is $\frac{\mu_{n}B}{k_BT_F}$, where $T_F$ is the
Fermi temperature and $\mu_n$ is the nuclear magnetron for $^3$He,
while for our calculation the dimensionless exchange splitting is
given by $\frac{V_{xc}}{W}$ where $W=16t$ is the bandwidth. The
experiments of Remeijer \etal were not performed at constant
pressure, which complicates the analysis somewhat, however they
conclude that

\begin{eqnarray}
\frac{T_C^{A_1}-T_C^{A_2}}{T_C^A} = \widetilde{a}
\left(\frac{\mu_{n}B}{k_BT_F} \right) + \widetilde{b}
\left(\frac{\mu_{n}B}{k_BT_F} \right)^2
\end{eqnarray}
where $\widetilde{a} = 36.3 \pm 0.91$ and $\widetilde{b} = 522 \pm
17$ in the range $0 \leq \frac{\mu_{n}B}{k_BT_F} \leq 0.01$ at an
effective pressure of 3.4~MPa i.e the splitting is, to a very good
approximation linear. The equivalent exchange splitting in our
calculations is $V_{xc} \le 0.01\,W = 0.01$~eV. It can clearly be
seen from figure \ref{fig:numerical_res_zz} that our calculations
give an approximately linear splitting between the A$_1$ and A$_2$
phase transitions over the range of exchange splitting $0 \leq
V_{xc} \leq 0.01$~eV. Hence our results are consistent with the
what is known about $^3$He. (Although, of course, we had no right
to expect this agreement as our parameters where chosen for \ZZ
and not $^3$He.) Further this illustrates the fact that
ferromagnetic superconductors will provide an excellent laboratory
in which to study the splitting of the A$_1$ and A$_2$ phase
transitions (and the non-linear splitting in particular) over a
far greater range of exchange splitting than is possible in
$^3$He. Further, when the effects of scattering from non-magnetic
impurities are included this model gives results that are
qualitatively consistent with the observed pressure dependence of
$T_C$ in \zz \cite{BenThesis,Ben2}.

\section{Discussion}

We have derived gap equations for superconductivity in coexistence
with ferromagnetism. We have done this for s-wave singlet states and for
p-wave triplet states with either ESP or OSP pairing. We used these gap
equations to study the behaviour of these states as a function of
exchange splitting.

For the singlet state we found that our gap equations reproduced
the Clogston--Chandrasekhar limiting behaviour and the phase
diagram of the Baltensperger--Sarma equation (neglecting the possibility of
an FFLO state).
We also showed that the singlet gap equation leads to
the result that the superconducting order parameter is independent
of exchange splitting at zero temperature. This fact was assumed in
the derivation of the Clogston--Chandrasekhar limit.

OSP triplet states showed a very similar behaviour to the singlet
state in the presence of exchange splitting. This leads to the
conclusion that the effect of exchange splitting on a
superconducting state is determined by whether the state contains
OSP or ESP. (All singlet states are, by definition, OSP states.)

In contrast, ESP triplet states show a very different behaviour in an exchange
field. In particular there is no Clogston--Chandrasekhar limiting.
Further, $T_C$ is actually increased by exchange splitting because
$D_\sigma(\varepsilon_F)$ is changed by the exchange splitting and
$T_C$ is dependent on $D_\sigma(\varepsilon_F)$. This effect is
well known in $^3$He, but has previously only been studied in a
Ginzburg--Landau formalism \cite{Brussaard}. The gap equations
presented here will allow for far more detailed study of both the
increase of $T_C$ and for the study of the splitting of the A$_1$
and A$_2$ phases by exchange splitting.

If the experimentally occurring ferromagnetic superconductors are
ESP triplet pairing states, as seems likely from the absence of
Clogston--Chandrasekhar limiting, then these systems will allow
for study of this effect at far greater exchange splittings than
can be archived with magnetic fields in $^3$He. The gap equations
presented here will also be useful for studying these materials in
their own right, in particular we hope that the will prove useful
for identifying the superconducting pairing symmetry of these
ferromagnetic superconductors. Our formalism is quite general, and
can be applied to more realistic band structures and pairing
models, although the additional complication of the vector
potential will have to be overcome before one can make complete
theoretical predictions for these materials.

\ack{We would like to thank the Laboratory for Advanced
Computation in the Mathematical Sciences
(http://lacms.maths.bris.ac.uk) for extensive use of their beowulf
facilities. One of the authors (BJP) was supported by an EPSRC
studentship and, in the final stages of this work, by the
Australian Research Council.}


\section*{References}
\begin{thebibliography}{99}
\bibitem{Saxena} Saxena S S, Agarwal P, Ahilan K,
Grosche F M, Haselwimmer R K W, Steiner M J, Pugh E, Walker I R,
Julian S R, Monthoux P, Lonzarich G G, Huxley A, Sheikin I,
Braithwaite D and Flouquet J 2000 \emph{Nature} {\bf 406} 587

\bibitem{URhGe}
Aoki D, Huxley A, Ressouche E, Braithwaite D, Flouquet J, Brison
J-P, Lhotel E and Paulsen C 2001 \emph{Nature} {\bf 413} 613

\bibitem{Pfleiderer}
Pfleiderer C, Uhlarz M, Hayden S M, Vollmer R, von~L{\"o}hneysen
H, Bernhoeft N R and Lonzarich G G 2001 \emph{Nature} {\bf 412} 58

\bibitem{Buzdin}
Buzdin A I and Bulaevskii L H 1986 \emph{Sov. Phys. Usp.} {\bf 29}
412 (1986 \emph{Usp. Fiz. Nauk.} {\bf 149} 45)

\bibitem{Bulut}
Bulut N, 2002 \emph{Adv. Phys.} {\bf 51} 1587

\bibitem{Eskildsen}
Eskildsen M R, Harada K, Gammel P L, Abrahamsen A B, Andersen N H,
Ernst G, Ramirez A P, Bishop D J, Mortensen K, Naugle D G,
Rathnayaka K D D and Canfield P C 1998 \emph{Nature} {\bf 393} 242

\bibitem{Mathur}
Mathur N D, Grosche F M, Julian S R, Walker I R, Freye D M,
Haselwimmer R K W and Lonzarich G G 1998 \emph{Nature} {\bf 394}
39

\bibitem{Ross_review}
McKenzie R H 1998 \emph{Comments Cond. Matt.} {\bf 18} 309

\bibitem{Fay&Appel}
Fay D and Appel J 1980 \PR B {\bf 22} 3173

\bibitem{Yates} Yates S
J C, Santi G, Hayden S M, Meeson P J and Dugdale S B \emph{Phys.
Rev. Lett.} 2003 {\bf 90} 057003

\bibitem{Cordes}
Cordes H G, Fischer K, and Pobell F 1981 \emph{Physica} B {\bf
107} 531

\bibitem{Deursen}
van~Deursen A P J, Schreurs L W M, Admiraal C B, de~Boer F R and
de~Vroomen A R 1986 \emph{J. Magn. Mater} {\bf 54-57} 1113

\bibitem{Balian&Werthamer}
Balian R and Werthamer N R 1963 \PR {\bf 131} 1553

\bibitem{Vollhardt}
Vollhardt D and W{\"o}lfe P 1990 \emph{The Superfluid Phases of
Helium 3} (London: Taylor and Francis)

\bibitem{Ben1}
Powell B J, Annett J F and Gy{\"o}rffy B L 2002 \emph{Ruthenate
and Rutheno-Cuprate Materials: Unconventional Superconductivity,
Magnetism and Quantum Phase Transitions (Springer Lecture Notes in
Physics vol~603)} ed Noce C, Cuoco M, Romano A and Vecchione A
(Heidelberg: Springer)

\bibitem{BenThesis}
Powell B J 2002 \emph{On the Interplay of Superconductivity and
Magnetism} Ph.D. thesis, Unversity of Bristol, UK

\bibitem{Fulde&Ferrell}
Fulde P and Ferrell R A 1964 \PR {\bf 135} A550

\bibitem{Larkin&Ovchinnikov}
Larkin A I and Ovchinnikov Yu N 1965 \emph{Sov. Phys. JETP} {\bf
20} 762

\bibitem{Mineev&Samokhin}
Mineev V P and Samokhin K V 1999 \emph{Introduction to
Unconventional Superconductivity} (Amsterdam: Gordan and Breach)

\bibitem{de_Gennes_SC}
de~Gennes P G 1966 \emph{Superconductivity of Metals and Alloys}
(New York: W.A.~Benjamin)

\bibitem{Ketterson&Song}
Ketterson J B and Song S N 1999 \emph{Superconductivity}
(Cambridge: Cambridge University Press)

\bibitem{GyorffyStauton&Stocks}
Gy{\"o}rffy B L, Staunton J B and Stocks G M 1991 \PR B {\bf 44}
5190

\bibitem{Ben4}
Powell B J, Annett J F and Gy{\"o}rffy B L \emph{The Freedericksz
transition in a quasi two dimensional p-wave superconductor}
\emph{Preprint}

\bibitem{Clogston}
Clogston A M 1962 \PRL {\bf 9} 266

\bibitem{Chandrasekhar}
Chandrasekhar B S 1962 \emph{Appl. Phys. Lett.} {\bf 1} 7

\bibitem{Zuo}
Zuo F, Brooks J S, McKenzie R H, Schlueter J A and Williams J M
2000 \PR B {\bf 61} 750

\bibitem{Baltensperger}
Baltensperger W 1958 \emph{Physica} {\bf 24} S153

\bibitem{Sarma}
Sarma G 1963 \emph{J. Phys. Chem. Solids} {\bf 24} 1029

\bibitem{Sigrist&Ueda}
Sigrist M and Ueda K 1991 \RMP {\bf 63} 239

\bibitem{Ben2}
Powell B J, Annett J F and Gy{\"o}rffy B L \emph{Preprint}
cond-mat/0301364

\bibitem{Remeijer}
Remeijer P, Roobol L P, Steel S C, Jochemsen R and Frossati G 1998
\emph{J. Low Temp. Phys.} {\bf 111} 119

\bibitem{Brussaard}
Brussaard P and Capel H W 1999 \emph{Physica A} {\bf 265} 370
\end{thebibliography}
\end{document}